\documentclass[aps,prl,twocolumn,reprint,secnumarabic,nobalancelastpage,amsmath,
amssymb,nofootinbib]{revtex4-2}

\usepackage{graphicx}
\usepackage{longtable}     
\usepackage{url}           
\usepackage{bm}            
\usepackage{color}

\begin{document}

\title{A mechanistic model for the asymmetric torque-speed relationships of a bacterial flagellar motor}

\author{Biswajit Das}

\affiliation{
  Beijing International Center for Mathematical Research (BICMR) and \\
  Biomedical Pioneering Innovation Center (BIOPIC), Peking University, Beijing 100871, China}

\author{Hao Ge}
\affiliation{
  Beijing International Center for Mathematical Research (BICMR) and \\
  Biomedical Pioneering Innovation Center (BIOPIC), Peking University, Beijing 100871, China}

\date{\today}

\begin{abstract}
A tiny bacterial flagellar motor rotates in both counter-clockwise (CCW)
and clockwise (CW) rotational directions. The most important
measurable characteristic of the flagellar motor is its
torque versus angular speed
relationship in CCW or CW modes, which is found to be non-symmetrical with
each other, and still, such a phenomenon
is not clearly understood.
Here, we explain this asymmetry through a mechanistic model based
on the detailed torque analysis for the rotation of the motor and
the revolutionary as well as spinning motion of the filament
and bead. We find
out that the asymmetry results from the conformational
changing of the hook due to rotational switching,
rather than any non-symmetric changes in the
potential of mean force generated by the stator-rotor interactions.
In CCW mode, when the hook remains bend and flexible, the revolution motion
predominates and the restoring
torque in this motion, originated due to drag, governs the shape of the torque-speed
curve. However, in CW mode, spinning motion dominates as the hook becomes
straight and rigid, and the linear torque-speed relation arises
due to the restoring torque for the drag corresponding to this motion.
Our study indicates the significant role of the hook's
conformational change upon the biological functions of the motor and paves
the way for further experimental exploration on the structural
origin of such asymmetry.
\end{abstract}

\maketitle

\begin{figure*}
\includegraphics{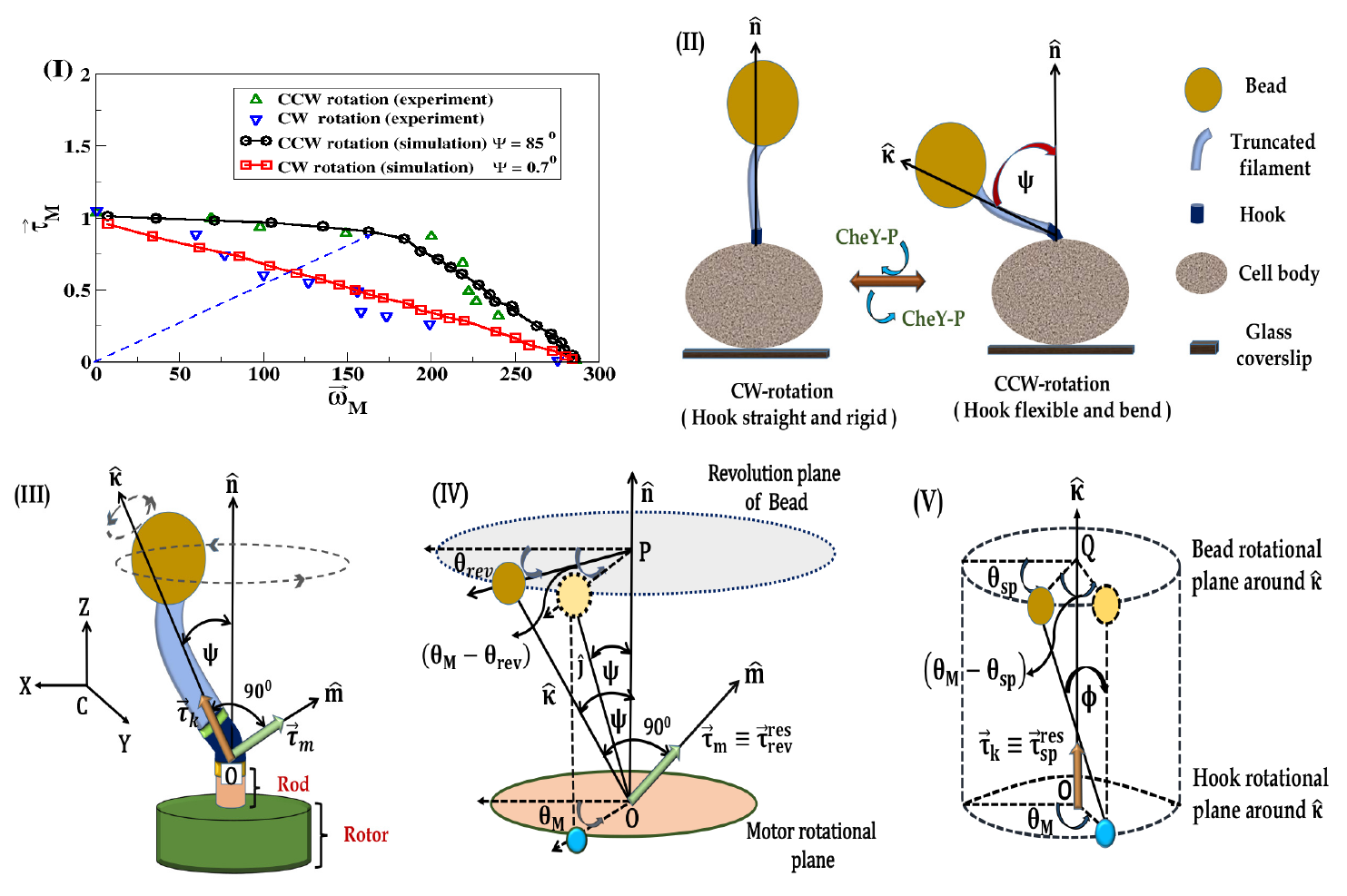}
\caption{\label{fig.1}(I) The
asymmetric torque-speed relations are numerically established for CCW
and CW rotations of the motor. Experimental data are considered from the work
of Yuan {\it et al.}\cite{Yuan-PNAS}.
(II) Significantly different bending angles in the CCW and CW modes.
(III) A general diagramatic
description of the origin of revolution and spinning motions
due to splitting of the motor torque, $\vec{\tau}_{R}$ into
$\vec{\tau}_{m}$ and $\vec{\tau}_{k}$ for bending of the hook
as an angle $\psi$. (IV) and (V) Schematic explanation of the generation of
the restoring torques during revolution and spinning motions
of FB, respectively.}
\end{figure*}

A flagellated bacteria, such as E. Coli, swims
in a run-tumble-run fashion
by modulating the orientation of the filaments from `bundle' to
`unbundle' phases\cite{Berg, Blair-FEBS-Lt,
Sowa, Darton-J-Bacteriology, Brown-J-Bacteriology, Wolde-MSB,
Fan-PRL, Relay-Nat-Sci-Rep, Block-Nature-Lett}.
Each filament is connected with a reversible motor
through hook\cite{Berg,Blair-FEBS-Lt,Sowa,Darton-J-Bacteriology,
Brown-J-Bacteriology, Wolde-MSB,Fan-PRL}
(see Fig.\ref{fig.1}).
When all the motors spin in the counter-clockwise(CCW) direction,
the filaments form a bundle
\cite{Berg,Blair-FEBS-Lt,Sowa,Darton-J-Bacteriology,
Brown-J-Bacteriology, Wolde-MSB},
whereas for the spinning of at least one motor in the clockwise(CW) direction,
filaments disentangle
\cite{Berg,Blair-FEBS-Lt,Sowa, Brown-J-Bacteriology,
Darton-J-Bacteriology, Wolde-MSB,Fan-PRL, Relay-Nat-Sci-Rep}.
The motor rotational switching
from CCW to CW direction, which is manifested by the attachment of
the chemotaxis signaling protein, CheY-P at the
switching or rotor complex of the motor
\cite{Paul-PNAS,Sarkar-PNAS,Blair-Biochemistry,Fan-Science}, can regulate
the movement of the cell.

The bacterial flagellar motor (BFM) is powered by the
flow of $H^{+}$ or $Na^{+}$-ions through the ion conducting
transmembrane channels in stator\cite{Blair-FEBS-Lt,Sowa}.
Morphologically, such $8$-$12$ non-rotating stators are placed at
the periphery of the rotor
\cite{Berg,Blair-FEBS-Lt,Sowa, Fan-Biophys-J,Xing-PNAS,Ryu-Nature,Meacci,
Gabel-PNAS,Lee-Nature,Turner-Biophys-J,Hosu-PNAS,Ryu-Nature,Chen-Biophys-J}.
During passing through the channel, the ions exert
the ion motive force on the stator
which is converted into the mechanical force by the stator for rotating
the rotor \cite{Xing-PNAS,Ryu-Nature,Meacci,Gabel-PNAS,Lee-Nature}. Consequently,
a torque is generated at the stator-rotor interface.

For elucidating the torque-generation(TG) mechanism as well as
the functional activity of motor,
experimentally, the relation between the motor torque, $\vec{\tau}_{M}$ and
the angular velocity of motor, $\vec{\omega}_{M}$ are measured
\cite{Ryu-Nature,Yuan-PNAS}.
In experiments, a polystyrene bead
is attached with the flagellar stub, and a weak optical
trap is used to measure the rotational speed of the bead which is modified
either by changing the bead size \cite{Xing-PNAS,Ryu-Nature} or
by regulating the viscocity of the medium\cite{Xing-PNAS,Chen-Biophys-J,Yuan-PNAS}.
For CCW rotation, the torque-speed (TS) relation is measured in absence
of  CheY-P; and  the TS curve  exhibits a plateau region,
where  $\vec{\tau}_{M}$ decreases slightly with $\vec{\omega}_{M}$ up
to an intermediate value, the `knee speed'
$(\approx 180 \hspace{0.1cm} rad/s)$, after
which it falls rapidly to zero (see experimental data of \cite{Yuan-PNAS}
depicted in Fig.\ref{fig.1}(I)
as green upper triangle). For CW rotation, the TS relation
was initially thought to be symmetric with that of the CCW rotation\cite{Xing-PNAS}.
However, recent experiment \cite{Yuan-PNAS} reveals that in presence of
the high concentration of CheY-P, there is no plateau region in the measured
TS relation where
torque decreases linearly with speed (see blue lower triangle in
Fig.\ref{fig.1}(I), taken from \cite{Yuan-PNAS}). Therefore,
an asymmetry is observed in the TS relations for CCW and
CW rotations of the motor, in absence and in
presence of CheY-P proteins, respectively. Such an asymmetry is thought
to be optimized for bacteria to search the chemical attractants
\cite{Yuan-PNAS}, but the underlying mechanism is still unknown.

In this letter, we provide a mechanistic
explanation of the asymmetric TS relations by noticing
that the motor rotation, regulated by the stator-rotor interactions, is also
modulated by the conformational dynamics of the hook.
A soft bending hook appears in CCW rotation
\cite{Brown-J-Bacteriology,Xing-PNAS,Relay-Nat-Sci-Rep},
responsible for showing the revolution motion of the
filament and bead (FB) predominantly, while in CW rotation,
hook changes itself into a straight rigid one
\cite{Darton-J-Bacteriology,Brown-J-Bacteriology,Relay-Nat-Sci-Rep}
(see Fig.(\ref{fig.1})(II)), liable for mainly the spinning motion of the FB.
The restoring torques developed due to drag acting on FB
in the revolution and spinning motions can modulate the motor torque in
different enormities; and consequently, the asymmetry in the torque-speed
curves is observed in CCW and CW rotational modes.

In our model, the most crucial parameter is the bending angle
of the hook, $\psi$ (Fig.(\ref{fig.1})(II)). We consider here that
in CW rotation, hook remains along the direction
normal to the cell body surface at the location of
the motor \cite{Relay-Nat-Sci-Rep}, represented
by the unit vector $\hat{n}$ in Fig.(\ref{fig.1})(II);
while in absence of CheY-P in CCW-rotation,
hook bends due to elasto-hydrodynamic instability\cite{Relay-Nat-Sci-Rep}
and makes an angle $\psi$ \cite{Brown-J-Bacteriology},
between the unit vector $\hat{k}$ along
the axis of the filament and $\hat{n}$ (Fig.(\ref{fig.1})(II)).
Moreover, following \cite{Relay-Nat-Sci-Rep}, a sharp transition
of the hook's stiffness with $\psi$ is contemplated during rotational
switching of the motor.

The bending of hook initiates an
additional complexity, {\it i.e.,} it induces two different
motions of FB, revolution
around $\hat{n}$ and spin around $\hat{k}$ \cite{Shimogonya-Sci-Rep}.
The origin of these
two motions is the torque upon the hook generated by the rotation
of the motor \cite{Hosu-PNAS}, which can be orthogonally decomposed
into $\vec{\tau}_{k}$ and $\vec{\tau}_{m}$ (Fig.\ref{fig.1}(III)).
The torque $\vec{\tau}_{k}$ along $\hat{k}$ are
responsible for the spinning motion of FB with angular velocity
$\vec{\omega}_{sp}$, whereas $\vec{\tau}_{m}$ along
$\hat{m}$ participates to revolve FB with angular
velocity $\vec{\omega}_{rev}$, where $\hat{k}$ and $\hat{m}$
are two orthogonal unit vectors(Fig.\ref{fig.1}(III)).

Denote $\vec{\tau}_{m}^{\hat{n}}=\vec{\tau}_{m}\sin\psi$ and
$\vec{\tau}_{k}^{\hat{n}}=\vec{\tau}_{k}\cos\psi$
as the components of $\vec{\tau}_{m}$ and $\vec{\tau}_{k}$,
respectively along the direction of $\hat{n}$, we can write down
the torque balance equation of motor:
\begin{equation}
 \zeta_{M}\vec{\omega}_{M}=\vec{\tau}_{SR}-\vec{\tau}_{m}^{\hat{n}}-\vec{\tau}_{k}^{\hat{n}}+ \sqrt{2k_{B}T\zeta_{M}}\xi_{M},
\label{Eq-motor}
\end{equation}
where $\vec{\tau}_{SR}$ is the rotational torque originated from
the stator-rotor interaction,
$\zeta_{M}$ and $\vec{\omega}_{M}$ are the rotational drag
coefficient and angular velocity of the motor, respectively; and
$\sqrt{2k_{B}T\zeta_{M}}\xi_{M}$ is the Brownian torque
acting on the motor with $k_{B}$, $T$ and $\xi_{M}$ are the Boltzmann constant,
absolute temperature, and the uncorrelated Brownian noise, respectively.

Similarly, the torque balance equation of FB for the revolution motion is
\begin{equation}
\zeta_{rev}^{B}\vec{\omega}_{rev}=\vec{\tau}_{m}^{\hat{n}}+
\sqrt{2k_{B}T\zeta_{rev}^{B}}\xi_{rev},
\label{Gen-eq-BF-rev}
\end{equation}
and for the spinning motion is
\begin{equation}
\zeta_{sp}^{B}\vec{\omega}_{sp}=\vec{\tau}_{k}+
\sqrt{2k_{B}T\zeta_{sp}^{B}}\xi_{sp},
\label{Gen-eq-BF-spin}
\end{equation}
where $\zeta_{rev}^{B}$
and $\zeta_{sp}^{B}$ are the drag coefficients of the FB for these
two motions, respectively (for more details, see \cite{SI}).
Similar to Eq.(\ref{Eq-motor}), here in
Eq.(\ref{Gen-eq-BF-rev})-(\ref{Gen-eq-BF-spin}), $\xi_{rev}$ and
$\xi_{sp}$ carry the same meaning as $\xi_{M}$.
$\vec{\omega}_{rev}$ and $\vec{\omega}_{sp}$ are the angular
velocities of FB for the revolution and spinning motions, respectively.

Once the hook behaves flexibly and bends, the hydrodynamic drag
acting on FB makes a delayed movement of it compared to the motor,
which is shown in Fig.\ref{fig.1}(IV) and (V) for the
revolution and spinning motions, respectively.
However, as FB is attached with the motor through hook, so
motor would endeavour
to rotate the FB along with the same speed of itself. Consequently,
due to elastic nature of the hook, the restoring torques,
$\vec{\tau}_{rev}^{res}$ and $\vec{\tau}_{sp}^{res}$
would be originated in both motions.  Neglecting the inertia of
the hook, the torque balance of the hook states
that for the revolution motion, $\vec{\tau}_{rev}^{res}\equiv \vec{\tau}_{m}$,
whereas $\vec{\tau}_{sp}^{res} \equiv \vec{\tau}_{k}$ for the spinning motion.
Hence we express $\vec{\tau}_{rev}^{res}$ and
$\vec{\tau}_{sp}^{res}$ as
\begin{equation}
\vec{\tau}_{rev}^{res}=\kappa_{h}(\theta_{M} - \theta_{rev}),
\label{Eq-res-torq-rev}
\end{equation}
and
\begin{equation}
\vec{\tau}_{sp}^{res}=\kappa_{sp}\left(\theta_{M}-\theta_{sp}\right),
\label{Eq-res-torq-spin}
\end{equation}
where $\theta_{M}$ designate the angular displacement of motor
at time $t$, and $\theta_{rev}$ as well as $\theta_{sp}$ convey the
same meaning
of FB for its revolution and spinning motions, respectively.
Here, $\kappa_{h}$ designates the
bending stiffness of hook in the XY-plane, whose value becomes
equal to the stiffness of hook
in the XZ-plane as hook is considered here as an elastic thin tube\cite{SI}.
For spinning or torsional rotation,
$\kappa_{sp}$ in Eq.(\ref{Eq-res-torq-spin}) represents the
torsional stiffness of hook,
\begin{equation}
\kappa_{sp}=\kappa_{h}/\lambda,
\label{relation-kh-ksp}
\end{equation}
where $\lambda=0.05$\cite{Flynn-Biophys-J-2004}.
Moreover, we consider the variation of $\kappa_{h}$ with $\psi$
according to the observation in \cite{Relay-Nat-Sci-Rep},
which is implemented here through a Hill-type equation,
\begin{equation}
\psi=\psi_{max}\left[1-\left\{\frac{1}
{1+\left(\frac{\kappa^{(sc)}_{h}}{\kappa_{h}}\right)^{\gamma}}\right\}\right],
\label{Eq-psi-k}
\end{equation}
where $\kappa^{(sc)}_{h}=960$ pN-nm/rad \cite{Relay-Nat-Sci-Rep}
represents the critical bending stiffness of hook and
two parameters, $\psi_{max}$ and $\gamma$ designate the
maximum bending angle and stiffness of the $\kappa_{h}$ versus
$\psi$ curve \cite{SI}. The result of
\cite{Relay-Nat-Sci-Rep} is incorporated through our
proposed equation (eq.(\ref{Eq-psi-k}))
by considering the fact that in CCW rotation of motor, hook remains flexible as
$\kappa_{h} < \kappa^{(sc)}_{h}$, which ensures the `run'
motion\cite{Relay-Nat-Sci-Rep}, whereas for CW rotation,
$\kappa_{h} > \kappa^{(sc)}_{h}$ resulting the tumble motion of
a bacterium
\cite{Brown-J-Bacteriology,Block-Nature-Lett,Relay-Nat-Sci-Rep}.

\begin{figure}
\includegraphics[width=0.45\textwidth]{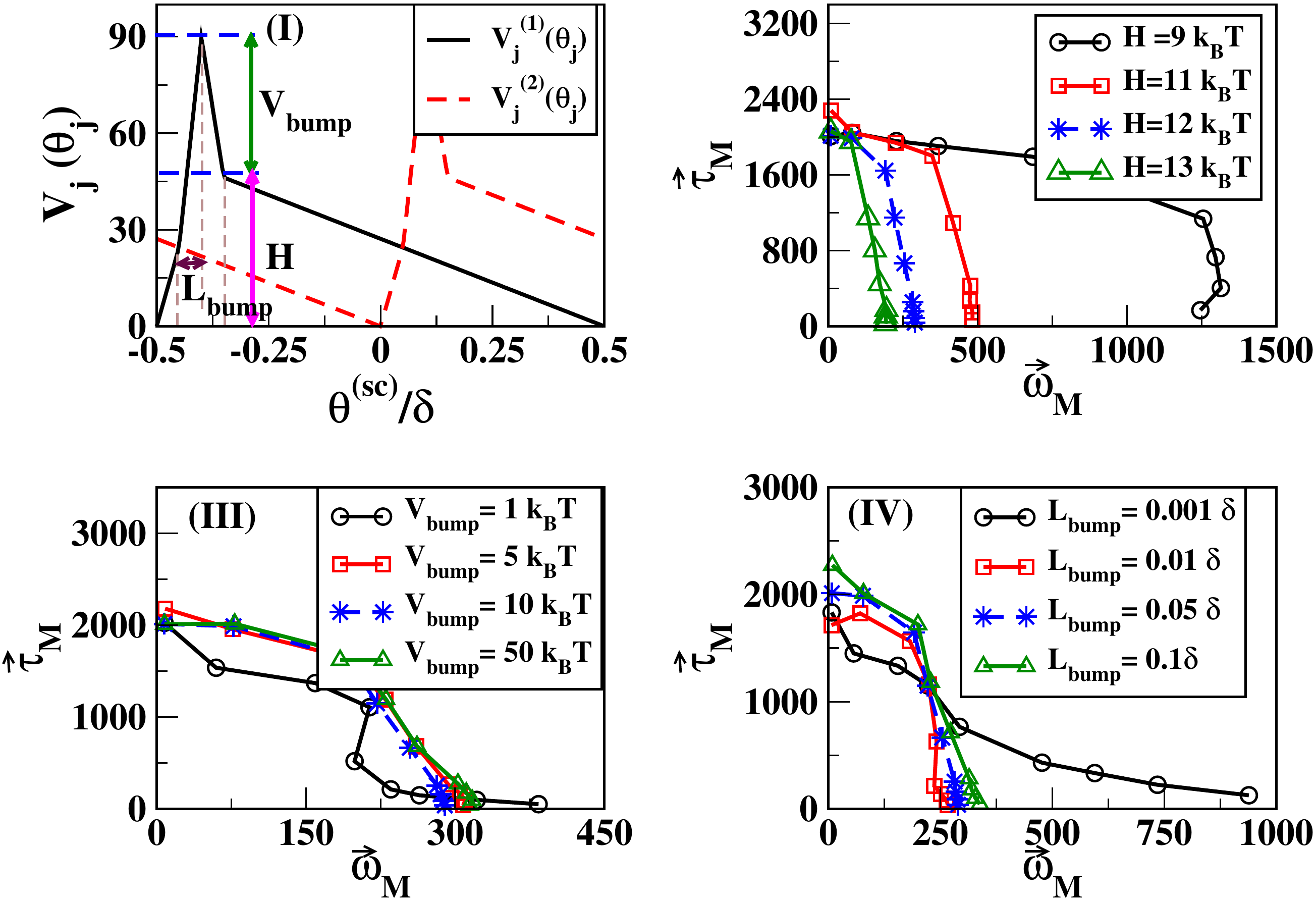}
\caption{(I) Forward
$V_{j}^{(1)}(\theta_{j})$ and $V_{j}^{(2)}(\theta_{j})$ are depicted,
where the meaning of $H$, $V_{bump}$, and  $L_{bump}$ are provided
graphically. $\theta^{(sc)}$
is the scaled rotational angle of
$\theta_{j}$ \cite{SI}.
Variation of $\vec{\tau}_{M}$ versus $\vec{\omega}_{M}$ for
the potential parameters $H$, $V_{bump}$, and  $L_{bump}$
are depicted in (II)-(IV), respectively for eleven load values, mentioned
details in \cite{SI}.}
\label{fig.2}
\end{figure}

In addition, in this study, we adopt the switching-diffusion type
TG mechanism
described in \cite{Xing-PNAS} for the estimation of $\vec{\tau}_{SR}$
in Eq.(\ref{Eq-motor}). Two asymmetric free-energy
 potentials $V_{j}^{(1)}(\theta_{j})$ and $V_{j}^{(2)}(\theta_{j})$,
with $V^{(2)}_{j}(\theta_{j})=V_{j}^{(1)}(\theta_{j}+0.5\delta)$
for $\delta = 2\pi/26$,
are considered to describe the interaction between
$\rm FliG$ proteins
with protonated and unprotonated conformations of the
$j$-th stator, respectively
(see Fig.{\ref{fig.2}}(I)). $\theta_{j}$ is defined as
$\theta_{j}=(\theta_{M}-\theta^{j}_{S})$,
where $\theta^{j}_{S}$ is the rotational angle of the $j$-th stator;
and the asymmetry in the potentials determine the direction of
rotation.
The torque $\vec{\tau}_{SR}  =-\sum_{j=1}^{N_{s}}\left(\partial V^{(i)}_{j}
(\theta_{j})/\partial \theta_{M}\right)$ with $i=1$ or $2$,
and $N_{s}$= 8 designates the total number of stators.
The conformational switching of stator is described as chemical
reactions, {\it i.e.}
$$V_{j}^{(1)}(\theta_{j})\underset{k_{-1}}{\stackrel{k_{1}}{\longleftrightarrow}}
V_{j}^{(2)}(\theta_{j}),$$
with transition rates
$$k_{\pm 1}= k^{0}_{\pm 1}exp\left[\mp\left(V_{j}^{(1)}(\theta_{j})- V_{j}^{(2)}(\theta_{j})\right)/2 k_{B}T\right],$$
where $k^{0}_{\pm 1}$ are the specific rate coefficients.
The potential and the kinetic parameters are adjusted in such
a way so that the transition
from $V_{j}^{(1)}(\theta_{j})$ to $V_{j}^{(2)}(\theta_{j})$ or vice-versa
follows the tight coupling mechanism between the
motor's rotation and the conformational switching of the stator
\cite{Xing-PNAS,SI}.

Finally, the averaged motor torque is calculated
at steady state in the form as
\begin{equation}
\vec{\tau}_{M}=\left(\zeta_{M}+\zeta_{B}\right) \langle \vec{\omega}_{M}\rangle,
\label{Eq-Gen-Torque-cal}
\end{equation}
where $\zeta_{B}=(\zeta_{rev}^{B}+\zeta_{sp}^{B}\cos\psi)$\cite{SI}. The average
value of $\vec{\omega}_{M}$,
$\langle \vec{\omega}_{M}\rangle $ is numerically calculated by solving
Eqs.(\ref{Eq-motor})-(\ref{Gen-eq-BF-spin}) combined with Eqs.
(\ref{Eq-res-torq-rev})-(\ref{Eq-psi-k}).
At first, the TS relation for CCW rotation is simulated with
$\kappa_{h}=1\times 10^{2}\hspace{0.1cm}{k_{B}T} <\kappa^{(sc)}_{h} $  for
$\psi\simeq 85^{0}$, calculated from Eq.(\ref{Eq-psi-k}); and
such a value of $\kappa_{h}$ is supported by
\cite{Block-Nature-Lett,Brown-J-Bacteriology}.
Our simulated TS curve makes a good agreement with
the experimentally determined curve \cite{Yuan-PNAS} (Fig.{\ref{fig.1}}(I)).

Next, we would investigate the rationale behind the linear
TS relation for CW rotation.
Logically, if the asymmetricity entirely depends on the TG mechanism,
the linear TS relation should be reproduced
through the variation of the parameters of the
potential of mean force, {\it i.e.} $V_{j}(\theta_{j})$.
To examine the possibility, a thorough study is carried out to understand
the variation of the nature of the TS relation
 with the parameters of the potential.
In Fig.(\ref{fig.2}) (II)-(IV), we depict the variation of TS relation for
three main parameters, {\it e.g.,} height of
$V_{j}(\theta_{j})$, $H$;
height of the bump of $V_{j}(\theta_{j})$, $V_{bump}$; and $L_{bump}$
 regulates the span of the bump region.
The thorough analysis is provided in \cite{SI},
and for clear understanding about the meaning of these three
parameters, see Fig.(\ref{fig.2})(a).
Our study reveals the fact that
the linear TS relation of CW rotation can never be obtained
from the variation of the parameters of $V_{j}(\theta_{j})$. 
Therefore, we can suggest that the asymmetry in the TS relations
for CCW and CW rotations does not solely depend on the TG mechanism.

From our simulation, the linear TS relation for CW rotation, described
in \cite{Yuan-PNAS}, can be nicely reproduced (see Fig.\ref{fig.1}(I))
for $\kappa_{h}=3.31\times10^{3}\hspace{0.1cm}k_{B}T$ for $\psi=0.7^{0}$,
which is calculated from Eq.(\ref{Eq-psi-k}).
Moreover, in Fig.{\ref {fig.1}}(I), it can be noticed that
the ratio of $\vec{\tau}_{M}/\vec{\omega}_{M}$
is same at both CCW and CW modes, which is
depicted as a dotted blue line, and consistent
with the experimental observation in \cite{Yuan-PNAS}.

\begin{figure}
\includegraphics[width=0.44\textwidth]{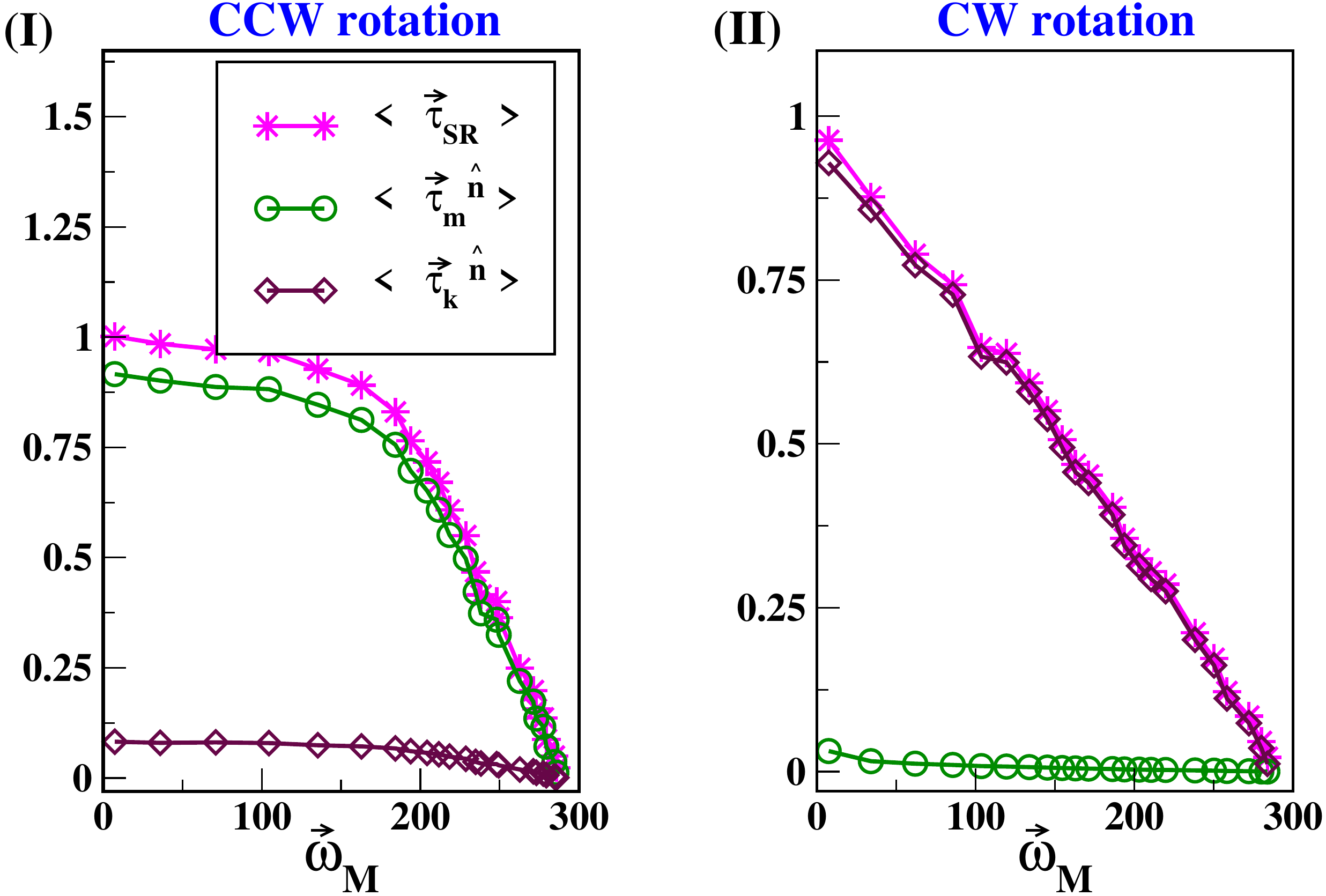}
\caption{(I) Normalized average values of $\langle\vec{\tau}_{SR}\rangle$,
$\langle\vec{\tau}_{m}^{\hat{n}}\rangle$ and
$\langle\vec{\tau}_{k}^{\hat{n}}\rangle$ with respect to the maximum value of $\langle\vec{\tau}_{SR}\rangle$ are plotted as a function of $\vec{\omega}_{M}$ for
CCW and CW rotations in (I) and (II), respectively.}
\label{fig.3}
\end{figure}

After next, we would scrutinize the reason for this asymmetry. Thus, in
Fig.\ref{fig.3}(I) and (II), the normalized value of
$\left\langle \vec{\tau}_{SR}\right\rangle\equiv \vec{\tau}_{M}$,
$\left\langle\vec{\tau}_{m}^{\hat{n}}\right\rangle$ and
$\left\langle\vec{\tau}_{k}^{\hat{n}}\right\rangle$ with respect to the maximum value of $\langle\vec{\tau}_{SR}\rangle$
are plotted as a function of $\vec{\omega}_{M}$ for CCW and CW
rotations, respectively. We observe that in CCW rotation (Fig.\ref{fig.3}(I)),
$\left\langle\vec{\tau}_{m}^{\hat{n}}\right\rangle$ varies
almost equally with $\left\langle \vec{\tau}_{SR}\right\rangle$,
whereas  $\left\langle\vec{\tau}_{k}^{\hat{n}}\right\rangle$
remains very small compare to
$\left\langle\vec{\tau}_{m}^{\hat{n}}\right\rangle$. Similarly,
in CW rotation, $\left\langle \vec{\tau}_{SR}\right\rangle$ and
$\left\langle\vec{\tau}_{k}^{\hat{n}}\right\rangle$ becomes linear
with $\vec{\omega}_{M}$ and varies almost equally. However, in this rotational mode,
$\left\langle\vec{\tau}_{m}^{\hat{n}}\right\rangle$ becomes negligible
 (see Fig.\ref{fig.3}(II)).

Since in Eq.(\ref{Eq-motor}), $\vec{\tau}_{m}^{\hat{n}}=\vec{\tau}_m\sin\psi$ and
$\vec{\tau}_{k}^{\hat{n}}=\vec{\tau}_k\cos\psi$, so when
 $\psi$ approaches to $90^{0}$ in the CCW rotation,
$\vec{\tau}_{k}^{\hat{n}}$ or
$\left\langle\vec{\tau}_{k}^{\hat{n}}\right\rangle$ is expected to be
negligible compared to
$\vec{\tau}_{m}^{\hat{n}}$ or
$\left\langle\vec{\tau}_{m}^{\hat{n}}\right\rangle$;
consequently, the revolution motion plays the most
significant role in determining
the nature of the TS curve. However, when $\psi$ approaches to $0^{0}$ in the
CW rotation, the revolution motion of FB tends to vanish as the 
radius of the
revolution plane goes to zero\cite{SI}. Thus, in the CW rotation,
only spinning motion of FB predominates, and $\vec{\tau}_{k}^{\hat{n}}$
or $\left\langle\vec{\tau}_{k}^{\hat{n}}\right\rangle$ mainly modulates
the motor torque.

In summary, in this letter, we provide a mechanistic explanation of
the experimentally determined asymmetric TS relations for
CCW and CW rotations of a BFM.
The bending of hook induces the revolution and spinning motions of FB;
and consequently, different restoring torques are developed for these two
motions due to hydrodynamic drag, which mainly modulate the
motor torque generated for the
stator-rotor interaction. We consider here that in presence of CheY-P proteins,
hook transforms its conformation from flexible to rigid one
by changing $\psi$ from $\simeq 90^{0}$ to $\simeq 0$ during
CCW to CW rotational switching of the motor, which is observed in
the in-vivo condition\cite{Brown-J-Bacteriology}. 
Only by tuning a single parameter, $\psi$,
our model can fit both of the experimentally determined TS relations of CCW
and CW rotations quite well.
The study reveals that in CCW rotation, revolution motion predominates,
and governs the
shape of the TS curve. However, in CW rotation, spinning motion
guides the shape of the linear TS relation.
Moreover, from our study, we can conclude that
the asymmetric TS relationship is not solely dependent on the TG mechanism;
rather it's a
consequence of the coupling between the torque-generation
mechanism and the conformational changing of the hook due to rotational
switching of a BFM.
Our study paves the way for further experimental exploration
on the structural origin of such asymmetry.

\begin{acknowledgments}
The authors acknowledge Prof. Fan Bai for his assistance at the
initial phase of this problem.  
H. Ge is supported by NSFC (Nos. 11622101 and 11971037).
\end{acknowledgments}

\end{document}